\documentclass[12pt]{iopart}

\usepackage{graphicx}

\usepackage{iopams}
\usepackage{setstack}

\begin{document}

\title[HALO enables ultra-low uncertainty measurements at kilowatts]{High amplification laser-pressure optic enables ultra-low uncertainty measurements at kilowatts}

%
\author{Alexandra B. Artusio-Glimpse, Kyle A. Rogers, Paul A. Williams, and John H. Lehman}

%
\address{National Institute of Standards and Technology, Boulder, CO, United States of America}

\ead{alexandra.artusio-glimpse@nist.gov}

\begin{abstract}
We present the first measurements of kilowatt laser power with an uncertainty less than 1~\%. These represent progress toward the most accurate measurements of laser power above 1~kW at 1070~nm wavelength and establish a more precise link between force metrology and laser power metrology. Radiation pressure, or photon momentum, is a relatively new method of non-destructively measuring laser power. We demonstrate how a multiple reflection optical system amplifies the pressure of a kilowatt class laser incoherently to improve the signal to noise ratio in a radiation pressure-based measurement. With 14 incoherent reflections of the laser, we measure a total uncertainty of 0.26~\% for an input power of 10~kW and 0.46~\% for an input power of 1~kW at the 95~\% confidence level. These measurements of absolute power are traceable to the SI kilogram and mark a state-of-the-art improvement in measurement precision by a factor of four.
\end{abstract}


\submitto{\MET}

\maketitle


\section{Introduction}\label{sec:intro}
We describe here our initial efforts to test the lower limits of measurement uncertainty for laser power using photon momentum as the measurement technique. In general, there are three approaches to the measurement of optical power distinguished by the mechanism of their response to light - thermal, electrical, and mechanical (photon momentum). In electrical approaches, light incident on a semiconducting photodiode or a photocathode generates an electrical current. The current indicates the incident optical power. Electrical approaches excel at high bandwidth measurements of laser power with extremely linear responsivities \cite{Zwinkels2010} and sensitivity deep into the ultraviolet \cite{Gottwald2018}. Thermal approaches equate incident optical power to the heat generated by the absorbed light. Their range encompasses powers from the single photon level \cite{Gerrits2019} to over 100~kW continuous wave (CW) \cite{Chamberlain1978,Schott1991,Day2005}. The lowest measurement uncertainty for optical power is achieved with a thermal technique implementing a cryogenic radiometer for an expanded relative uncertainty of 0.0052~\% \cite{Stock2000}. The third category of optical power measurement is through power meters that react the momentum from incident light and measure laser power as a force (photon momentum radiometers). This approach enables a paradigm-changing power measurement that no longer requires the light to be absorbed in order to be measured, it can merely be reflected from a mirror attached to a force sensor. This ``non-exclusive" approach allows not only simultaneous measurement and use of optical power but dramatically increases the upper limit of measurable power \cite{Williams2017b,Williams2021} by minimizing heating that is intrinsic to the absorbing techniques.

A key to low measurement uncertainty is achieving a high signal-to-noise ratio (SNR). The impressively low measurement uncertainty of the cryogenic radiometer comes from the noise reduction obtained with cryogenic cooling. In contrast, we seek here to increase SNR in photon momentum power measurement by increasing the signal. That is, we take advantage of the non-exclusive nature of radiation pressure to passively amplify the optical power. By making the laser beam reflect from the sensing mirror multiple times, we multiply its force on the mirror without a commensurate increase in noise. This technique offers the intriguing and unique possibility among high accuracy laser power metrology of making the lowest uncertainty power measurement at the highest possible power levels.

This passive amplification approach was demonstrated in radiation pressure-based power measurements first by Stimler \cite{Stimler1964} where pulsed laser light (2.5 J/pulse) reflected sequentially from two mirrors on opposing arms of a torsion balance; no uncertainty assessment was included. Later, Vasilyan et al. \cite{Vasilyan2017} and Manske et al. \cite{Manske2019} discussed potential reflection geometries and demonstrated up to 21 reflections of $\sim$1~W incident laser power with an estimated uncertainty of 10~\%. Shaw et al. also amplified $\sim$1~W of laser power with 7 reflections to yield an expanded relative uncertainty of 4~\% \cite{Shaw2019}. Even more recently, Vasilyan et al. \cite{Vasilyan2021} demonstrated up to 33 reflections of laser powers between 1 and 10~W using two sensing mirrors achieving a 4.6~\% (k=2) uncertainty at $\sim$8~W which included a 10~dB noise reduction due to the common-mode rejection by the dual sensing setup.

For high-accuracy metrology, these amplification schemes using incoherent (non-overlapping) reflections are attractive because their effective gain can be accurately quantified from the countable number of reflections, degraded by the finite reflectivity of the mirrors and system geometric factors. Significantly larger power amplifications on the order of 1000 \cite{Schwartz2017} can be attained when the optical power is circulated in a coherent cavity. When one of the cavity mirrors is a force sensor, such a device becomes a power meter with dramatic passive signal gain. But, this enhancement comes at the cost of significantly greater uncertainty on the gain factor. Wagner et al. \cite{Wagner2018} demonstrated a radiation pressure power measurement at 370 $\mu$W with a coherent circulating cavity yielding a gain factor of 250. Although the radiation pressure-based measurement of circulating power in the cavity had only a 3~\% measurement variability, the agreement with a force prediction based on cavity gain was only 20~\%.

For this reason, we have chosen an incoherent gain approach. We demonstrate low measurement uncertainty by measuring a large 10~kW optical power signal with 14 reflections from the sensing mirror using a force sensor with low environmental noise.

\section{Methods}\label{sec:methods}
\subsection{System design}\label{sec:systemDesign}
Our ultimate interest is to measure a 10~kW laser with an expanded relative uncertainty of 0.01~\%. This uncertainty level is on the same order as that achievable with cryogenic radiometer measurements of laser power at the sub-milliwatt level \cite{Gentile1996,Stock2000}. Toward this goal, we have designed a multi-reflection optical system that amplifies the radiation pressure force by a factor of 14. We call this optical system the High Amplification Laser-pressure Optic (HALO). A particular advantage of the HALO design is its compatibility with many different force sensors including the commercial force balance used in the presented measurements. Helping us toward our prospective goal of 0.01~\% relative uncertainty, the HALO is also compatible with electromagnetic force balances, like those developed at NIST \cite{Shaw2019b,Keck2021}, which are traceable to electrical SI units and obviate imprecise mass artifact calibration (see \sref{sec:measurementUncertainty} for a discussion of uncertainty sources). A second requirement of the HALO architecture is maintaining a consistent laser angle of incidence on the high-reflectivity mirror that receives the radiation pressure force. This protects the mirror from thermal flexing and excessive optical loss, because the high reflectivity coating need only be optimized for a single angle of incidence. Fixing the incident angle also simplifies the propagation of uncertainties. In this System design section, we outline the HALO system architecture, its construction, and detail the measurement gains expected.

\begin{figure}[ht]
    \centering
    \includegraphics[width=0.8\textwidth]{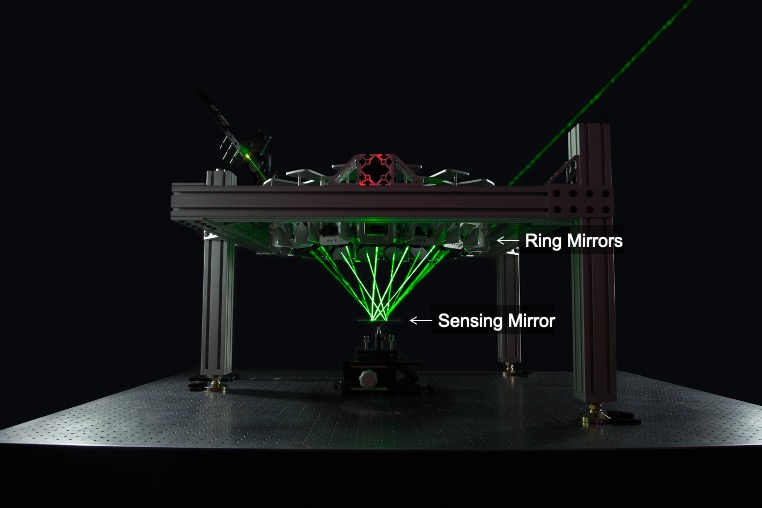}
    \caption{Photograph of the HALO system illuminated with a 5~mW green helium-neon laser (entering at left) to show the beam path in place of the high power infrared laser.}
    \label{fig:haloPhoto}
\end{figure}

\subsubsection{Architecture}\label{sec:architecture}
\begin{figure}[ht]
    \centering
    \includegraphics[width=1\textwidth]{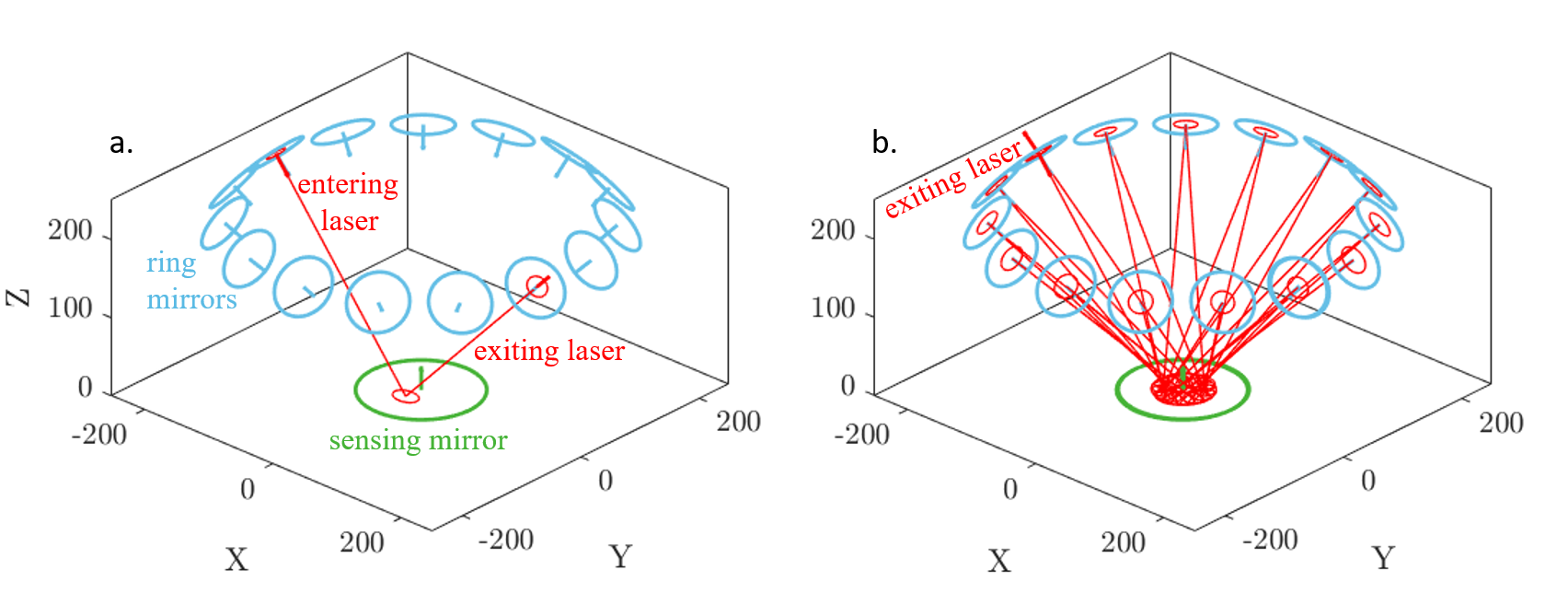}\label{fig:1-15bounce}
    \caption{Ray drawing of the HALO with a) 1 reflection off the sensing mirror and b) 15 reflections off the sensing mirror showing the path of a laser as it rotates through the system.}
\end{figure}

HALO is a pentadecagonal structure with an entrance port and 15 ring mirror modules, 13 of which are filled with mirrors that redirect the input laser beam back toward sensing mirror, always at an incident angle of 45$^\circ$. In \fref{fig:haloPhoto}, a green 5~mW helium-neon laser illuminates the beam path through the system. The beam enters the system from the upper left side of the photograph and exits the system to the upper right of the image. A ring of mirrors, individually referred to as the ring mirrors, fold the beam back toward the sensing mirror, centrally located toward the bottom of the image, to build up radiation pressure gain. In this image, 13 ring mirrors are used to achieve a gain of 14, but any number of reflections from 1 to 15 (\fref{fig:1-15bounce}(a) and \ref{fig:1-15bounce}(b), respectively) are possible with this structure by adding or removing the appropriate ring mirrors from the modules.

\begin{figure}[ht]
    \centering
    \includegraphics[width=0.6\textwidth]{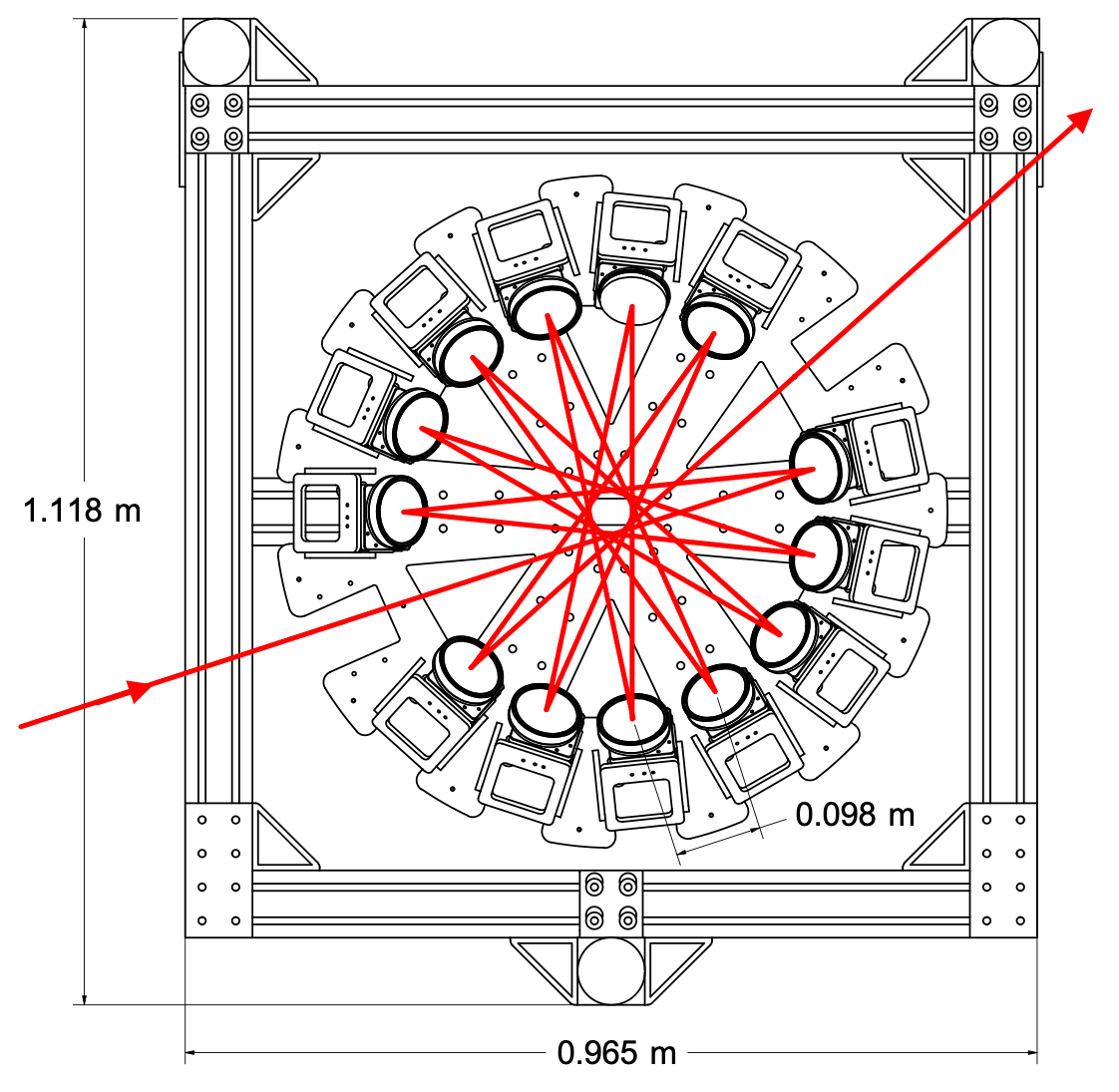}
    \caption{Ray path viewed from perspective of force sensor projected onto the plane of ring mirrors with the ring mirror structure drawn in the background. Projected onto this two-dimensional plane, the star pattern design is much like a toroidal multipass absorption cell.}
    \label{fig:starPattern}
\end{figure}

Viewed from the perspective of the force sensor in \fref{fig:starPattern}, we see the star pattern that the laser beam follows, drawn for 14 bounces. This design is much like toroidal multipass cells for laser spectroscopy \cite{Tuzson2013}; however, here the center of the star pattern is moved out of the plane of the ring mirrors. The sensing mirror is placed at this out-of-plane star pattern center. Like the toroidal cells, the modular ring mirrors of the HALO system could be replaced with a continuous ring mirror with an entrance and exit aperture. A continuous ring mirror would have the benefit of simplifying the process of changing the number of reflections, by simply changing the laser entrance angle, and would decrease the number of alignment degrees of freedom. However, such a large, precision optic would have high expense.

To accommodate a $\sim$25~mm diameter, 10~kW, 1070~nm wavelength laser beam with a divergence angle of 1~mrad, we designed the HALO with 76.2~mm diameter, fused silica ring mirrors (each with 3-axis adjustment control) and a 150~mm diameter, fused silica sensing mirror. The entire structure fills a volume of 1.2~m$^3$ and is contained within a large air-shielding box (2.3~m$^3$ containing volume). Each of the ring mirrors is tilted downward out of the ring plane by 45.16$^\circ$ and the laser incidence angle at these mirrors is 4.24$^\circ$ with respect to each ring mirror surface normal. The height between the ring plane and the sensing mirror is 238.6~mm. The path length of the laser propagating from one ring mirror to the next is 674.9~mm. After 14 reflections off the sensing mirror, the laser beam travels 9.448~m. In other words, a 1~mrad diverging beam will expand by more than 18~mm in diameter by the time it exits the HALO system. See Supplemental Material A for detailed calculations of this geometry.

\begin{figure}[ht]
    \centering
    \includegraphics[width=0.5\textwidth]{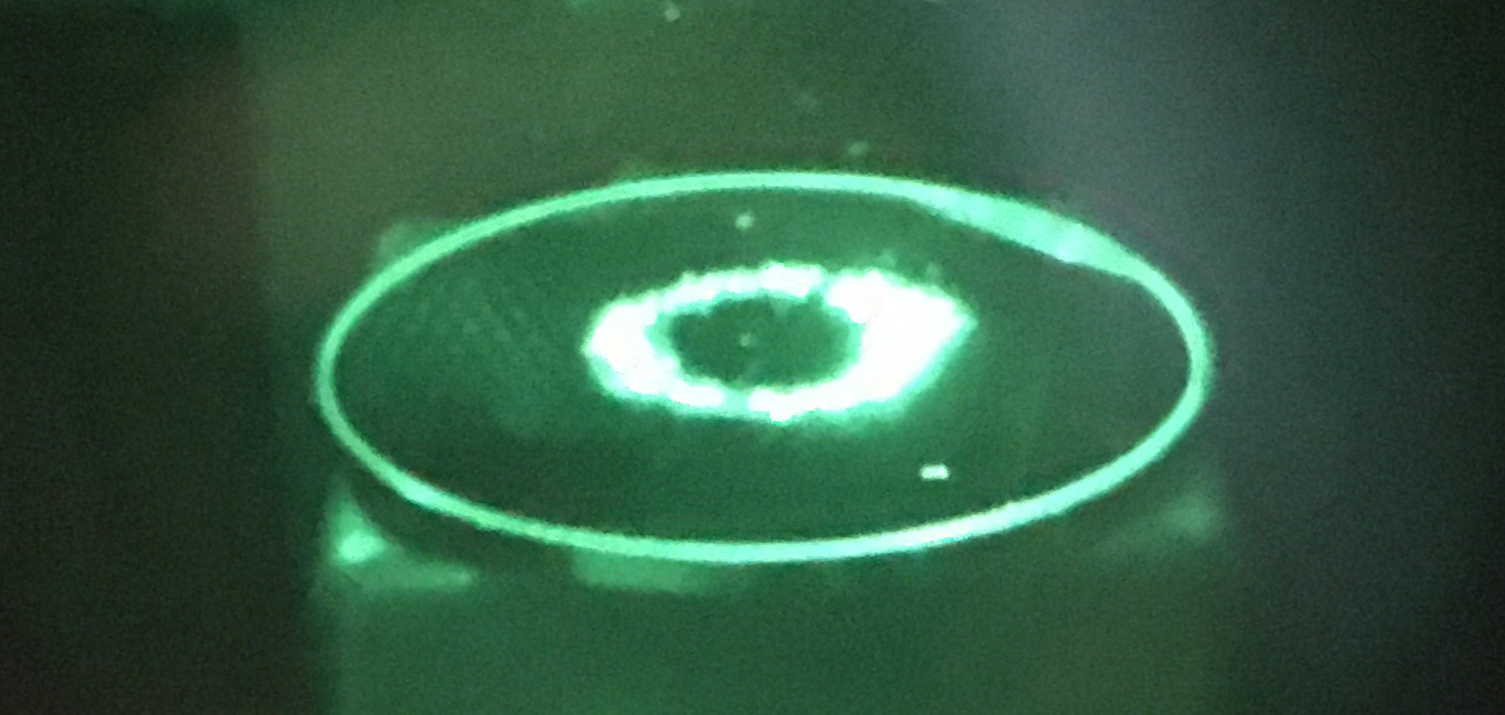}
    \caption{Photograph of the spot pattern of high-power infrared laser light scattering off the highly reflective sensing mirror and captured with an infrared camera. Variations in scattering angle change the spot intensity seen by the camera and do not represent a change in laser power on the mirror.}
    \label{fig:spotPattern}
\end{figure}

The laser spot pattern on the sensing mirror forms a ring with growing spot sizes as the laser rotates and expands through the system. Each spot center is approximately 10~mm from the next and positioned 25~mm from the center of the sensing mirror, deviating only by small misalignment errors. The regularity of this spot pattern may be used for sensitive feedback of the system optical alignment. \Fref{fig:spotPattern} is a photograph of the laser spot pattern on the sensing mirror seen because of small amounts of scattered light off the mirror surface.

\subsubsection{Construction}\label{sec:construction}
The HALO construction consists of two main components: the ring structure and the frame. The ring structure consists of two ring plates, the lower of which specifies the mount positions for each ring mirror and the upper of which adds additional rigidity. The frame provides support to the ring structure at a definable height. The full system is designed to sit atop an optical bench and provide accessibility while still allowing some disassembly for minor portability. Aluminum was chosen for the ring plate and mirror mounts for its high machinability and high strength to weight ratio.

The ring plates are light-weight yet stiff enough to minimize sagging under gravity. They each are made of an aluminum circular plate (76.2~cm in diameter and 6.35~mm thick) that are waterjet cut with symmetric, internal cutouts to reduce mass and provide the required access to each ring mirror assembly. These cutouts also allow transmitted light to pass out of the system. By designing two thin ring plates spaced by an extruded aluminum strut as opposed to a single, thick aluminum plate, we can minimize differential sagging of the plates under gravity. We achieve this with expandable spacers that force the two plates apart, stiffening the system. This stiffening also leads to an increase in the natural frequency of the structure, which is desirable because the force sensor is insensitive to frequencies above 10~Hz.

\begin{figure}[t]
    \centering
    \includegraphics[width=1\textwidth]{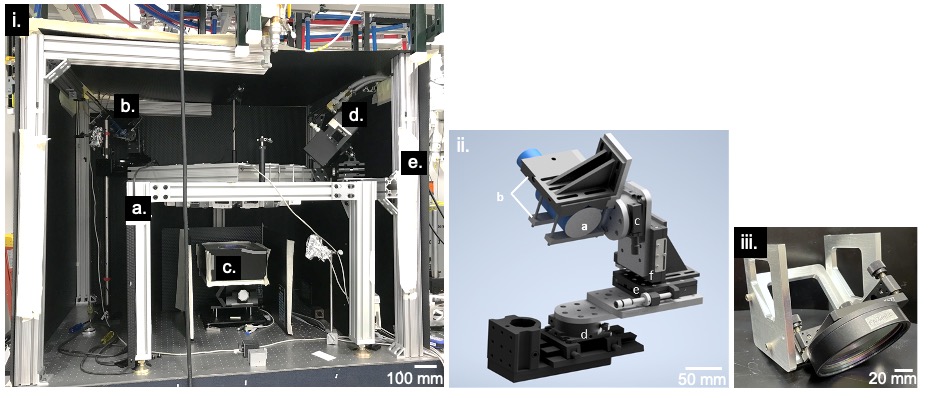}
    \caption{(i) Lab photograph of full system showing a) HALO structure, b) laser collimator mount, c) force sensor, and d) beam dump are all contained within e) a large, laser-safe and air current damping enclosure, which is pictured with one panel removed for an inside view. (ii) Laser collimator mount with the following components: a) laser collimator, b) v-block mount with hold-downs, c) tilt control stage, d) rotation control stage, e) horizontal adjustment stage, f) vertical adjustment stage. (iii) The ring mirror assembly fixes the angle of each ring mirror at 45.16$^\circ$, while a 3-axis kinematic mount affords fine scale tip, tilt, and z-position adjustability.}
    \label{fig:construction}
\end{figure}

Each mirror module is another light-weighted aluminum plate assembly. The assembly consists of three aluminum plates bolted together, where the central plate forms a 45.16$^\circ$ angle with respect to the top face of the assembly that bolts to the ring plate. These assemblies are interchangeable between ring mirror ports. A kinematic optical mount carries the 76.2~mm diameter ring mirror while providing tip, tilt, and z adjustability. \Fref{fig:construction}(iii) shows a closeup of one of the ring mirror assembly structures with the mirror in place.

The HALO frame is visible in \fref{fig:construction}(i) - a. Aluminum T-slotted frames with $50\times50$~mm cross-sectional profile offers high strength to weight ratio, providing adequate support for the two ring plates and all ring mirrors. A combination of flat-plate fasteners and corner gussets are used between beams to allow for a reasonably large ring mass. A three-legged design allows for coarse leveling of the ring mirrors with respect to the force sensor using height-adjustable feet. More accurate alignment is achieved by leveling the force balance (\fref{fig:construction}(i) - c). Lastly, the ring assembly can move vertically along the frame legs for coarse height adjustment up to 600~mm to accommodate force sensors of different heights. 

Another essential component of the system is the laser collimator mount. In order to achieve the necessary entrance angle of the laser light, a mount with multiple adjustable axes is necessary. The laser collimator has a diameter of 50~mm, so we make use of a clamped v-block. The v-block is bolted to two rotation stages, controlling the azimuth and elevation of the collimator. The mount also allows for fine positioning in the vertical and horizontal directions using micrometer-driven stages. Very precise alignment of the laser beam entering the ring is essential, as small misalignments of the laser will propagate through the system. A diagram detailing the individual parts of the mount is shown in \fref{fig:construction}(ii) and a photograph of the mount with the HALO system is given in \fref{fig:construction}(i) - b.

The last component that is not part of the HALO itself but is still critical for functionality is the enclosure, see \fref{fig:construction}(i) - e, where the front face of the enclosure has been removed to expose the components inside. The enclosure serves two main functions: air current damping and laser safety. Because the HALO utilizes a precision force balance with a large area mirror acting as the force receiver, air currents present when measuring laser power can cause significant measurement noise. To reduce this noise, a secondary internal enclosure is positioned directly around the force sensor, acting as an air baffle. Due to the high CW laser powers propagating out of plane of the optical table, all enclosure materials are of high laser-absorbing quality to limit damage to the system and fully contain stray light from escaping into the laboratory space. Aluminum frames support each laser safety panel rated for 1,200~W/cm$^2$ for up to 100~seconds, and the panels can be easily removed for accessibility.

\subsection{Theory}\label{sec:theory}
A high-power, nearly collimated laser reflected by an ultra-high reflectivity mirror imparts a force to the mirror along the mirror surface normal of
\begin{equation} \label{eq:power2force}
    F_1 = P/c \left[ (1-T+R)\cos(\theta)\cos^2(\psi/2) + SB_f \right]
\end{equation}
where $c$ is the speed of light in vacuum, $P$ is the incident laser power, $T$ is the transmittance of the mirror, $R$ is the reflectance of the mirror, $S$ is the non-specular scattering fraction of the mirror, $\theta$ is the angle of incidence of the laser relative to the surface normal of the mirror, $\psi$ is the divergence angle of the beam, and $T+R+S+A = 1$, where $A$ is the fraction of absorbed light. Nominal values for each of these terms are given in \tref{tbl:terms}. The shape of the diffuse scatter ($S$) off the surface of the mirror is described by the coefficient $B_f$. If the mirror surface is Lambertian this coefficient equals 2/3 or if the scatter is totally specular this coefficient equals 1 \cite{McInnes2004}. In the measurements reported herein, the mirror substrate is 3~mm thick fused silica coated on both faces with an ultra-high reflectivity coating such that $R\sim0.999987$. This reflectance is calculated from knowledge of the other three parameters: measured transmittance, calculated absorptance from material properties, and estimated scatter from a comparable measurement. An identically coated fused silica substrate, smaller in size and therefore compatible with the scattering measurement system, was measured at the NIST Bidirectional Optical Scattering Facility \cite{Germer1999} and used to estimate the scatter coefficient $S\sim0.4\times10^{-6}$ and the non-Lambertian coefficient $B_f\sim0.99994$. Keeping only the terms that contribute more than $10\times10^{-6}$ to the force, the reduced version of \eref{eq:power2force} is
\begin{equation} \label{eq:power2forceReduced}
    F_1 = P/c (1-T+R)\cos(\theta).
\end{equation}
The subscript ``1" is used to denote that this is the force transferred when the laser reflects off the mirror only once. To determine the force imparted to the mirror by the second reflection of the laser, we must keep track of any lost light in the round trip from the first reflection to the second reflection. If the folding optics have non-zero fractional optical loss, $L$, then the force imparted to the mirror by the laser’s second reflection off its surface is
\begin{equation} \label{eq:power2force2}
    F_2 = (1-L) R \left[P/c (1-T+R)\cos(\theta)\right],
\end{equation}
where accounted for are both the optical loss of the folding optics and the reduced power left after the first interaction with the sensing mirror. The $N$th reflection of the laser off the mirror imparts the following force, if all folding optics are identical and the angle of incidence with the primary mirror is constant:
\begin{equation} \label{eq:power2forceN}
    F_n = \left((1-L)R\right)^{N-1} F_1.
\end{equation}
Thus, the total force imparted to the mirror by $N$ reflections of the laser is
\begin{equation} \label{eq:power2forceMulti}
    F = F_1 \sum_{j=1}^{j=N} \left((1-L)R\right)^{j-1}.
\end{equation}
In the limit as $L\rightarrow0$ and $R\rightarrow1$, the summation approaches $N$.

\fulltable{\label{tbl:terms}Summary of terms in Equations~(\ref{eq:power2force}-\ref{eq:power2forceMulti}).}
        \br
        Equation & Symbol & Description & Nominal value \\
        \mr
        \ref{eq:power2force},\ref{eq:power2forceReduced} & $P$ & Incident laser power & 500 - 10000~W \\
        \ref{eq:power2force},\ref{eq:power2forceReduced} & $c$ & Speed of light & 299,792,458~m/s \\
        \ref{eq:power2force},\ref{eq:power2forceReduced} & $\theta$ & Angle of incidence & 45$^\circ$ \\
        \ref{eq:power2force} & $\phi$ & Laser divergence angle & 1~mrad \\
        \ref{eq:power2force},\ref{eq:power2forceReduced} & $T$ & Transmittance of sensing mirror & $7\times10^{-6}$ \\
        \ref{eq:power2force},\ref{eq:power2forceReduced},\ref{eq:power2forceMulti} & $R$ & Specular reflectance of sensing mirror & $\sim$0.999987 \\
        \ref{eq:power2force} & $S$ & Diffuse scatter of sensing mirror front surface & $\sim0.4\times10^{-6}$ \\
        \ref{eq:power2force} & $B_f$ & Front surface scatter non-Lambertian coefficient & $\sim$0.99994 \\
        \ref{eq:power2force} & $A$ & Absorptance of mirror & $\leq5\times10^{-6}$ \\
        \ref{eq:power2forceMulti} & $L$ & Loss of folding mirrors & $\sim8.8\times10^{-6}$ \\
        \ref{eq:power2forceMulti} & $N$ & Number of laser reflections off sensing mirror & 14 \\
        \br
\endfulltable

Our HALO system consists of 8~mm thick, coated fused silica ring mirrors. On the front side of the substrate, the mirrors are coated with a high-reflectivity ion beam sputtered coating, and on the back side, they are coated with an antireflective coating. The measured transmittance of these mirrors is $2\times10^{-6}$. Based on well-known material properties of the fused silica substrate and of the absorptivity of similar ion beam sputtered low-loss coatings, we estimate the absorptance of each ring mirror is $\leq5\times10^{-6}$. Over each round trip, the laser passes through 675~mm of 50~\% humidity air. Given the wavelength, 1070~nm, and full width half max bandwidth, 4~nm, of this laser, the estimated free-path absorption of the laser as it propagates through air in one round trip is $1.4\times10^{-6}$. By estimating that the scatter off these mirrors is the same as that of the sensing mirror at $0.4\times10^{-6}$, we predict the total optical loss between each reflection off the sensing mirror is approximately $8.8\times10^{-6}$.

A sensor that measures the total force acting on the sensing mirror is subject to noise that is both dependent and independent on the amount of laser light in the system. We describe the noise in units of force on the sensor as
\begin{equation} \label{eq:noise}
    \eta = \eta_0 + \eta_{P}(P) + \eta_{N}(NP)
\end{equation}
where $\eta_0$ is the noise level when the laser is off, $\eta_{P}(P)$ is the power-dependent noise only related to the input power from the laser collimator (also nearly equal to the power incident on the beam dump seen in \fref{fig:construction}(i)-d), and $\eta_{N}$ is the amplification-dependent noise taken to be driven by not only the input laser power but also the number of reflections off the force sensor, where $\eta_{N} \propto P\sum_{j=1}^N ((1-L)R)^{j-1}$.

If we define the signal to noise ratio of this measurement as $F/\eta$, then
\begin{equation} \label{eq:snr}
    \frac{F}{\eta} = F_1 \sum_{j=1}^{j=N} \frac{((1-L)R)^{j-1}}{\eta_0+\eta_P(P)+\eta_N(NP)}.
\end{equation}
Once again, it is informative to look in the limit as $L\rightarrow0$ and $R\rightarrow1$:
\begin{equation} \label{eq:snrLimit}
    \frac{F}{\eta} \rightarrow \frac{NF_1}{\eta_0+\eta_P(P)+N\eta_N(P)}.
\end{equation}
In effect, both the cold noise ($\eta_0$) and the input power-dependent noise ($\eta_P$) are reduced by $N$, while the amplification-dependent noise ($\eta_N$) scales with N and, therefore, will not contribute to improvements in SNR.

\subsection{Data analysis}\label{sec:dataAnalysis}
In the next section, measurements of laser power with the HALO using a commercial force balance are reported. To obtain a single measurement of laser power, a baseline reading from the force balance is first acquired. Then, the laser is turned on, kept on for 30~s, and turned off. The response time of the force balance is orders of magnitude slower than the time it takes the laser to fully turn on. As such, the laser is effectively an instantaneous signal. Unless otherwise noted, a chain of 10 of these square-wave pulses are collected and analyzed using the so called ``AB" method of drift correction described by Swanson and Schlamminger in 2010 \cite{Swanson2010}. This analysis returns a single, averaged measurement of laser power ($P_i$) from the 10-injection series and the uncertainty of this value ($\sigma_{P_i}$) is given by the square root of the estimated variance of the mean (Eq. 21-22 \cite{Swanson2010}). We then repeat the measurement and analysis several times to obtain multiple independent measurements of laser power for a given setup and report the average of these ($\bar{P}$) with the reduced uncertainty given by
\begin{equation} \label{eq:meanError}
    \sigma_{\bar{P}} = \sqrt{ (\sigma_{P_1}/m)^2 + (\sigma_{P_2}/m)^2 + \cdots + (\sigma_{P_m}/m)^2  }
\end{equation}
for a set of $m$ repeat runs. In this way, the HALO system is evaluated as a detector of laser power over a range of input power levels from 500~W to 10~kW and the associated measurement uncertainties calculated.

\section{Results}\label{sec:results}
\subsection{Preliminary measurements}\label{sec:prelimMeasurements}
To prepare for the high-power measurements, we first evaluate the effective optical wavefront flatness of the HALO to determine how much added curvature is transferred to the beam as it propagates through the system. We also measure the Allan deviation of the system when the laser is off and when it is on to determine the optimal integration time in the face of heat driven noise and drifting.

\subsubsection{Beam divergence measurement and estimation of mirror curvature}\label{sec:curvature}
As discussed in \sref{sec:architecture}, the laser propagates 9.448~m through the HALO system. In this process, it interacts with the sensing mirror 14 times and each of the 13 ring mirrors once. The ring mirrors are all 8~mm thick, 76.2~mm diameter fused silica, which are thick enough to resist any significant sagging or bowing under gravity or from the mounting. However, the sensing mirror is 150~mm in diameter and only 3~mm thick fused silica. Either due to asymmetries in the webbed aluminum mount supporting the sensing mirror or residual stress from the mirror coatings, curvature of the sensing mirror can be significant and negatively impact the laser propagation through the HALO system. Curvature of the sensing mirror, or of the ring mirrors, leads to astigmatic aberrations in the laser if not corrected, causing clipping at worst and collection difficulties at the point of the beam dump at best. Therefore, we measured the aggregated wavefront curvature imparted by the HALO optics to a probe laser before injecting a high-power laser through the system by measuring the probe profile at different propagation distances.

To measure the distortion caused by the HALO system, we used an infrared probe laser with a 1.4~mm diameter defined at the $1/e^2$ Gaussian width. We first determined the divergence angle by measuring its beam profile at a distance of 406~mm and~2295 mm from the collimator output. By the growth in beam width, we determined the beam divergence to be (333~$\pm$~7)~$\mu$rad ($\pm$ here denoting the difference between vertical and horizontal measurements), which roughly agrees with our expectation. We then propagated the same laser through the HALO system and measured its beam profile at the exit port. Importantly, we find that the astigmatic aberration of the probe laser upon exiting the HALO system is small – a measured ellipticity of 84~\% as compared to 95~\% at the input of the HALO system. This is an indication that the wavefront error imparted to the laser by the HALO optics is minimal. By the measured divergence angle of the probe laser, we would expect the beam width ($1/e^2$) at the HALO exit port to be 3.3~mm. The beam width we measure at the HALO exit port is 5~mm. This difference of (1.7~$\pm$~0.5)~mm corresponds to an added divergence of (0.18~$\pm$~0.05)~mrad. Following Gaussian beam propagation in the far field, this added divergence is equivalent to an additional accumulated wavefront curvature of approximately -0.1~m$^{-1}$. Such wavefront curvature is not large, allowing us to conclude that the sensing mirror is reasonably flat.

\subsubsection{Allan deviation}\label{sec:allanDeviation}
Heating within the HALO system perturbs the measurement by adding slow moving drift as the various components, specifically the force sensor, heat and flex. In addition, heating of components, like the beam dump, drive convective air flow within the system causing additional turbulent noise. Thermal noise is a common challenge in radiation pressure measurements. In the HALO measurements, we incorporate physical baffles throughout the system, as described in \sref{sec:construction}, to shield the force sensor from excessive radiative heating and to inhibit large convective air currents, especially near the force sensor. These defensive measures, however, do not entirely eliminate the thermal effects. The measurements series must be designed with these considerations in mind. 

\begin{figure}[b]
    \centering
    \includegraphics[width=0.45\textwidth]{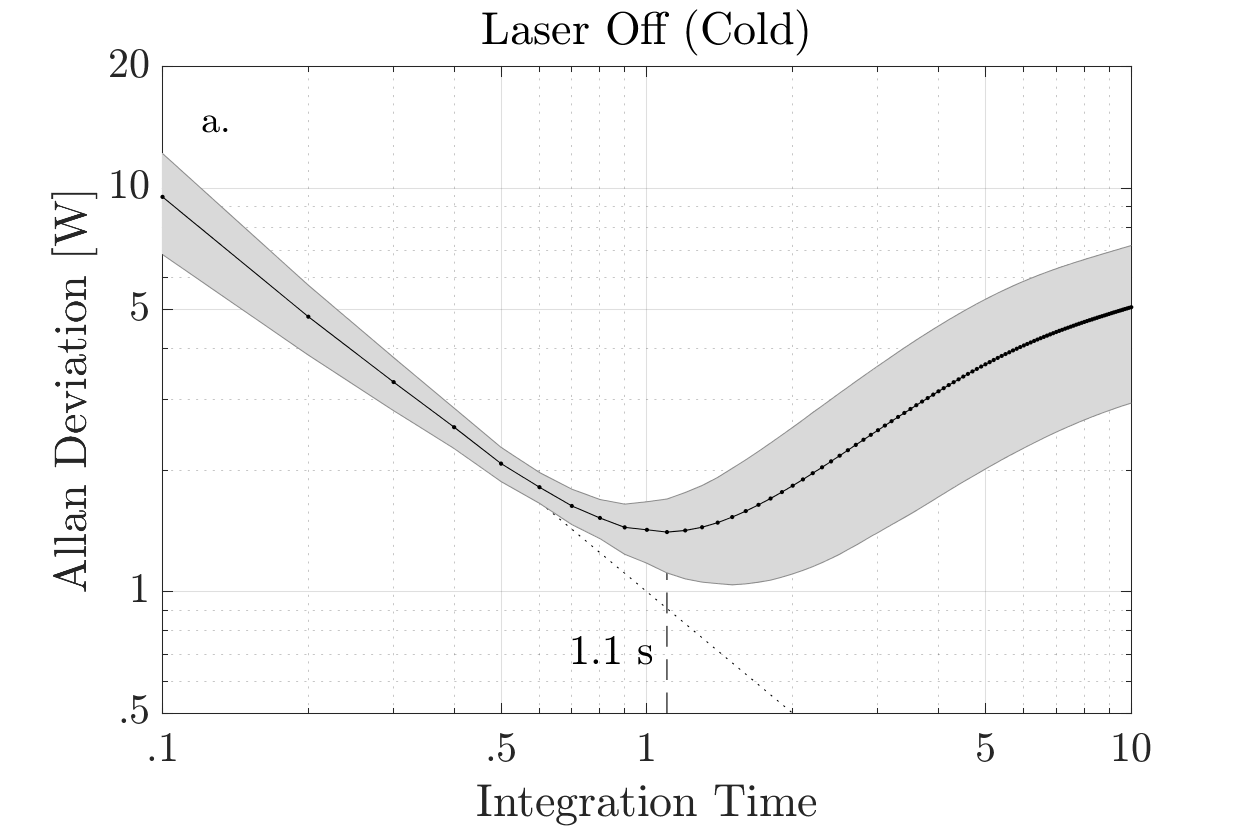}
    \includegraphics[width=0.45\textwidth]{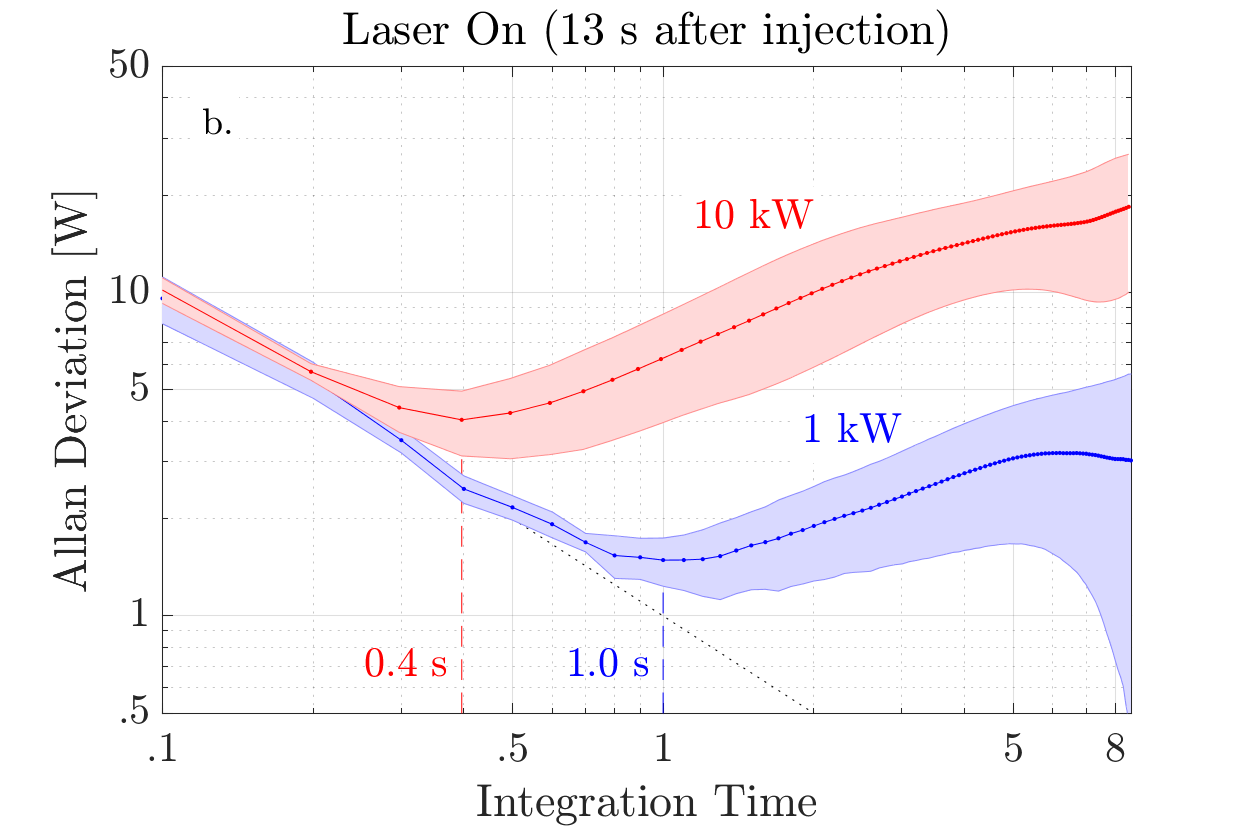}
    \caption{Allan deviation of the force balance response (in units of watts) (a.) when the laser is off and the system is cold and (b.) while the laser is on with 14 reflections and either 10~kW or 1~kW nominal injected laser power. Initially, these measurements are dominated by Gaussian white noise, but after 1~s (or 0.4~s for the high power data series), random walk noise dominates. }
    \label{fig:allanDeviation}
\end{figure}

To determine the laser injection duration that benefits most from averaging while minimizing thermal errors, we calculate the Allan deviation of the measured force signal in watts. A plot of this deviation as a function of integration time for a cold system (when the laser is off) and a hot system (when the laser is on) with 14 reflections off the sensing mirror and a nominal injected power of 1~kW and 10~kW is given in \fref{fig:allanDeviation}. Shaded areas in these plots indicate the standard deviation of several similar sets of data. For the ``laser on'' measurements, the integration begins 13~s after the laser is turned on, to exclude the slow response of the force balance to the step function signal of the laser \cite{Williams2021}. In these plots, the Allan deviation decreases to a minimum after 1~s (or 0.4~s in the case of the high power injection) before increasing again. This minimum point marks the balance between reduction of signal variation by averaging and increases to signal variation due to slow air current noise and thermal drifting. In all measurements of laser power, we selected a laser injection time of 30~s with a 35~\% duty cycle to allow the system ample time to cool before injecting the next laser pulse. We excluded the first 13~s of the force balance response to each laser injection and averaged the response for 1~s while the laser was on. 

\subsection{Measurements of laser power}\label{sec:measurementsOfPower}
\begin{figure}[ht]
    \centering
    \includegraphics[width=0.8\textwidth]{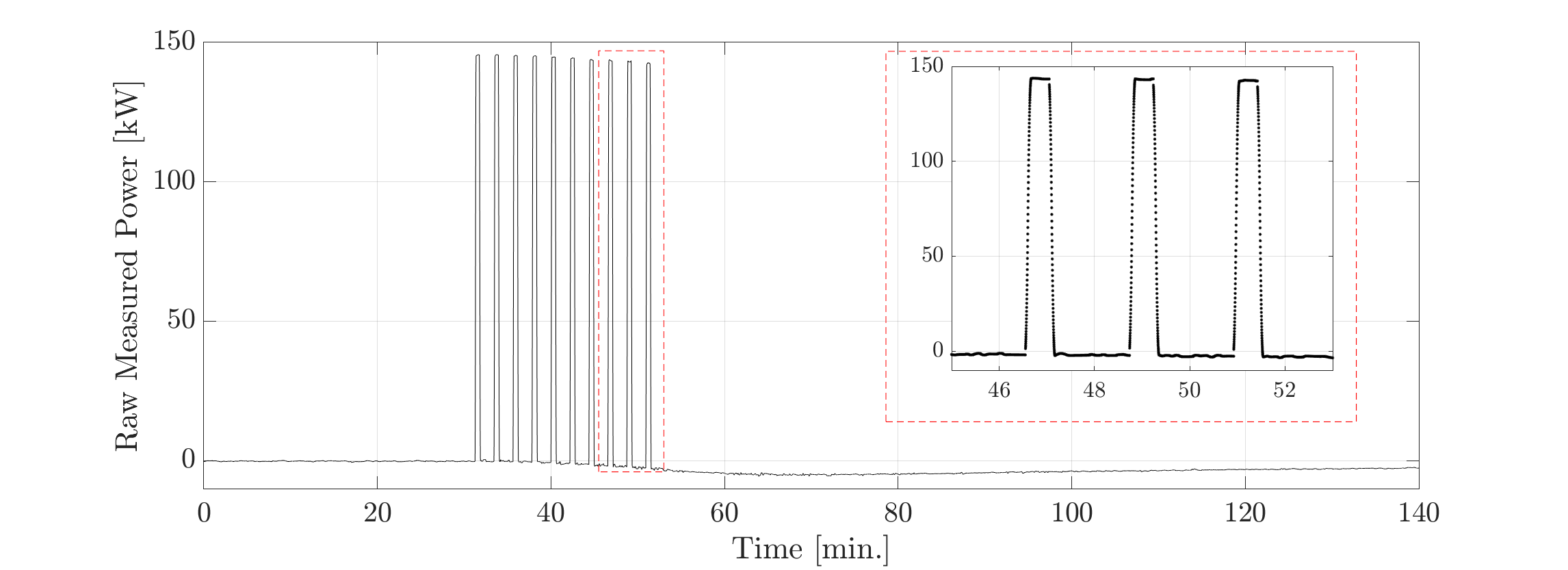}
    \caption{Uncorrected measurement traces of power measured with the commercial force balance showing response time of the sensor (inset) and baseline drifting. The baseline begins to drift $\sim$10~minutes after the series begins, reaching a maximum offset of -5~kW $\sim$10~minutes after the series ends.}
    \label{fig:rawPower}
\end{figure}

As discussed in \sref{sec:dataAnalysis}, we collect a series of 10 square-wave laser injections to analyze and extract a single measure of laser power. An example of such an injection series is given in \fref{fig:rawPower} for a nominally 10~kW laser injection. Notice that the measured power is about 145~kW due to the signal amplification of the 14-reflection system. To determine the raw measured power, the force balance measurement output in grams has been scaled by using \eref{eq:power2force}, where $F = mg$ and $g = 9.796022$~m/s$^2$~\cite{Pavlis2017}. Slow drifting over the series of injections is small but clear, especially just following the injection series where the drifting baseline drops to -5 kW before slowly returning to zero. This slow, nonlinear drifting is corrected in the data analysis \cite{Swanson2010}.

\begin{figure}[ht]
    \centering
    \includegraphics[width=0.45\textwidth]{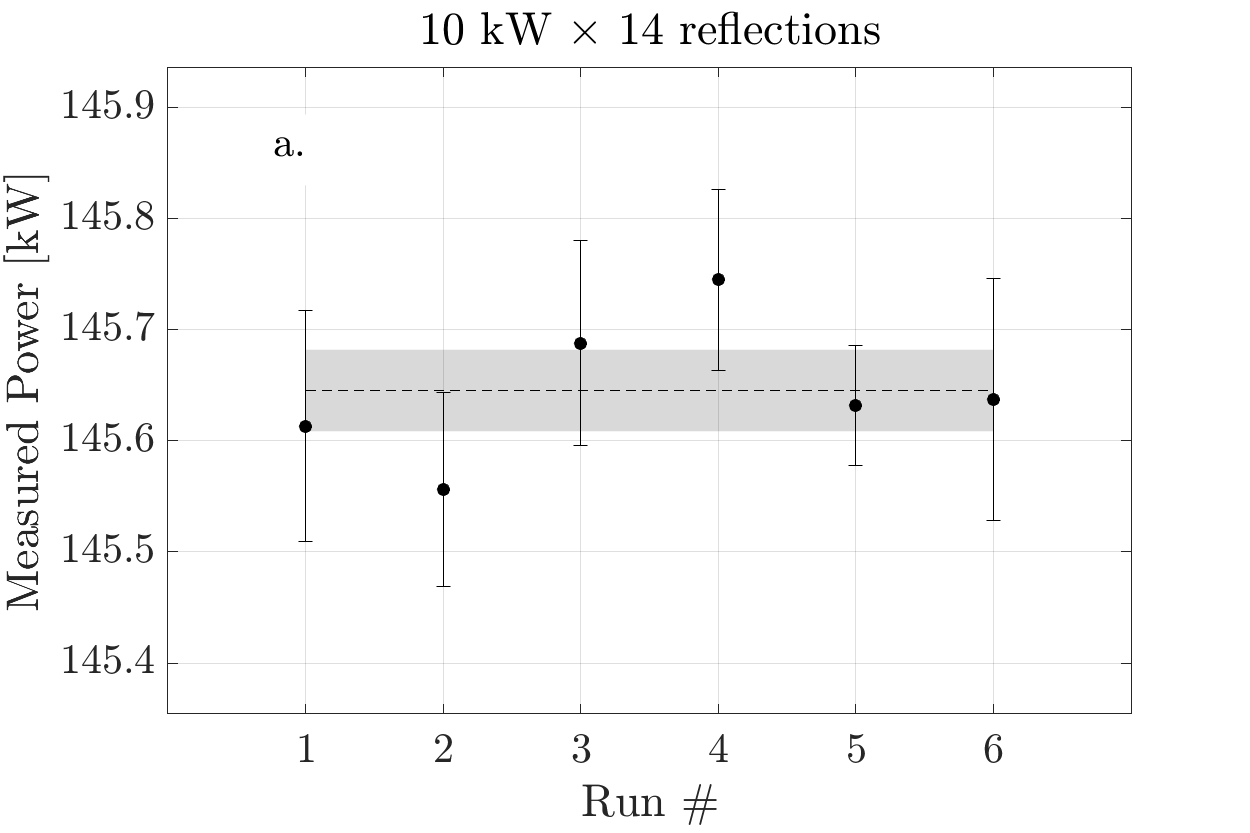}
    \includegraphics[width=0.45\textwidth]{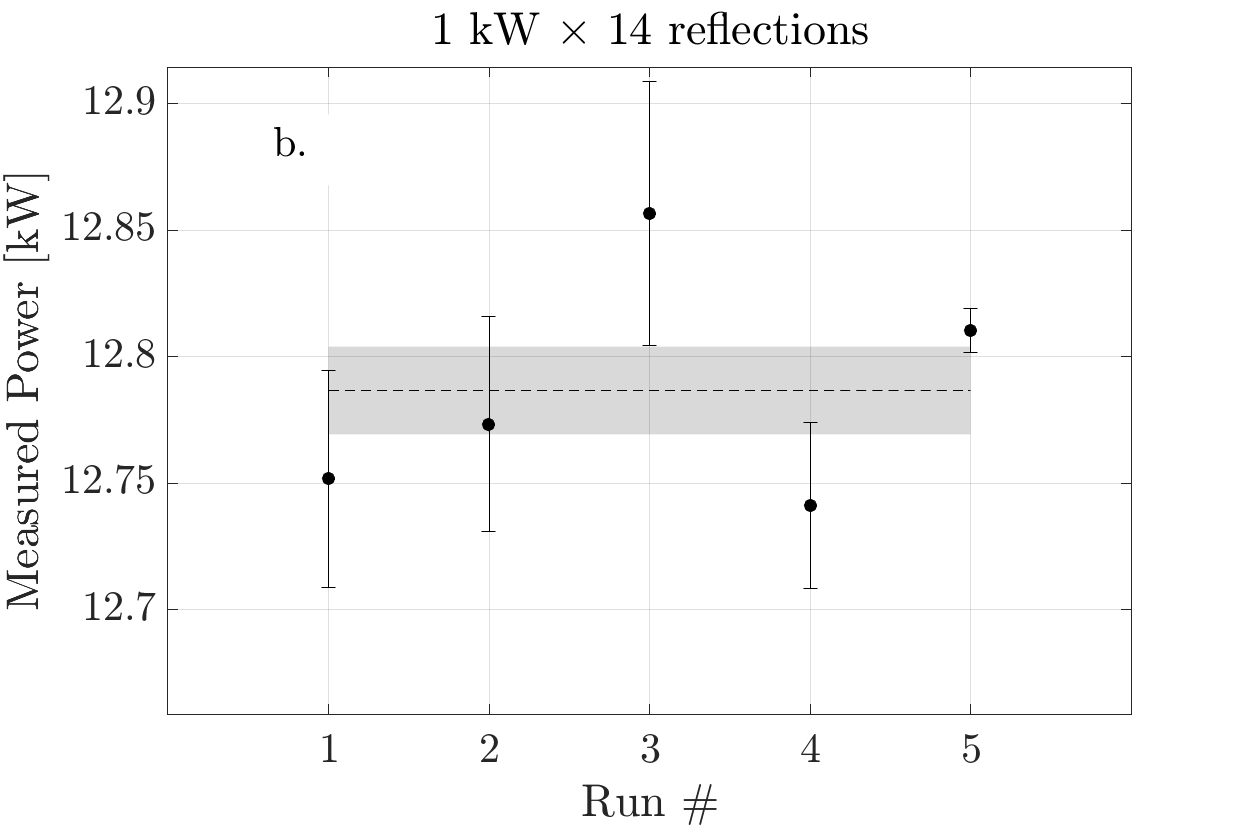}
    \caption{Corrected measurements of power from several injection runs, using the analysis method discussed in \sref{sec:dataAnalysis}. (a.) 10~kW laser input with 14 reflections repeated over 6 runs resulted in an average power of $\bar{P}=145.65$~kW (dashed line) and statistical uncertainty $\sigma_{\bar{P}} = 37$~W (gray region). (b.) 1~kW laser input with 14 reflections repeated over 5 runs resulted in an average power of $\bar{P}=12.79$~kW (dashed line) and statistical uncertainty $\sigma_{\bar{P}} = 17$~W (gray region). Error bars denote the square root of the estimated variance of each run.}
    \label{fig:processedPower}
\end{figure}

Following the AB drift correction calculation~\cite{Swanson2010}, we obtain a single corrected and averaged measurement of laser power for each 10-injection series. This procedure is then repeated multiple times to improve our measurement precision by averaging independent runs. \Fref{fig:processedPower} shows the results from multiple runs collected for 10~kW and 1~kW laser injected power. Note that these injected power levels are estimates used for labeling and are not precisely the injected laser power (which is unknown without directly measuring). By taking the average of these sets of runs, we obtain our final estimates of the measured laser power ($\bar{P}$). The error bars in \fref{fig:processedPower} are the square root of the estimated variance of each run value ($\sigma_{P_i}$) and the grey region gives the uncertainty of $\bar{P}$ ($\sigma_{\bar{P}}$ \eref{eq:meanError}).

\Tref{tbl:powerMeasurements} reports the final measured power ($\bar{P}$) for a range of injected power levels from 500~W to 10~kW. In two cases, at 1~kW and 10~kW, we also measure the response when the laser reflects off the sensing mirror only once (\fref{fig:1-15bounce}(a)) by removing the first ring mirror. These two measurements have an estimated gain factor of 1, while the measurements with the full HALO mirror system have an estimated gain factor of 14. The statistical uncertainties in \tref{tbl:powerMeasurements} are the square root of the estimated variance from each power measurement run, propagated through the mean calculation \eref{eq:meanError}.

\begin{table}
    \caption{\label{tbl:powerMeasurements}Power measurements using the HALO system. Nominal input power is the target laser output power; however, each run varied slightly from this value.}
    \begin{indented}
    \item[]\begin{tabular}{@{}lllll}
        \br
        Nominal & Measured & Number of & Number of & Statistical \\
        Input Power & Power ($\bar{P}$) & Reflections & Runs & Uncertainty ($\sigma_{\bar{P}}$) \\
        \mr
        10~kW & 145.65~kW & 14 & 6 & 37~W \\
        7~kW & 102.10~kW & 14 & 2 & 44~W \\
        1~kW & 12.79~kW & 14 & 5 & 17~W \\
        500~W & 7.20~kW & 14 & 2 & 20~W \\
        10~kW & 10.51~kW & 1 & 4 & 23~W \\
        1~kW & 0.94~kW & 1 & 5 & 23~W \\
        \br
    \end{tabular}
    \end{indented}
\end{table}

\subsection{Linearity and noise}\label{sec:linearityAndNoise}
With the set of power measurements from the HALO system given in \tref{tbl:powerMeasurements}, we study the power linearity of the system. We also estimate the measured amplification factor of implementing 14 reflections with respect to 1 reflection and analyze the measurement noise as given by $\sigma_{\bar{P}}$ to examine the amplification of noise within these measurements as the number of reflections is increased.

\subsubsection{Power linearity}\label{sec:powerLinearity}
\begin{figure}[ht]
    \centering
    \includegraphics[width=0.4\textwidth]{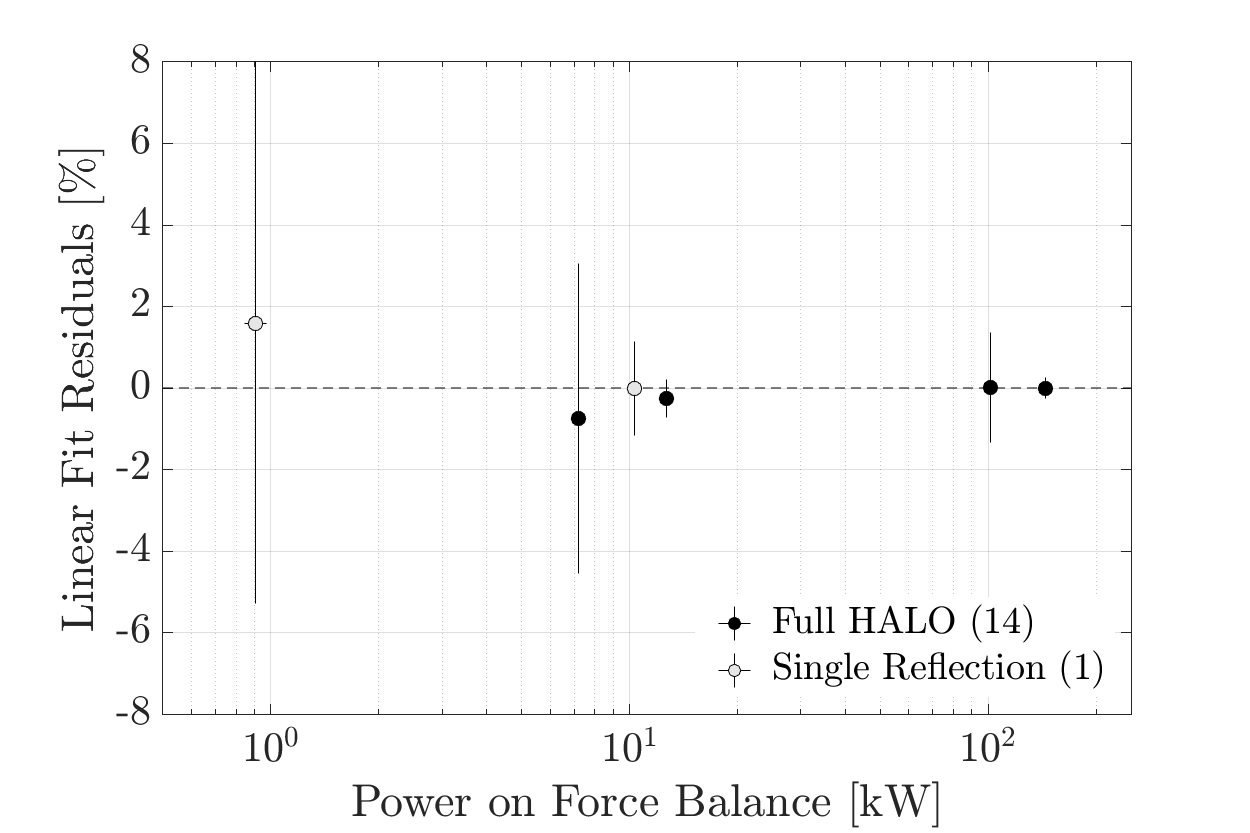}
    \caption{Linearity plot of HALO power measurement. Fit residuals are from comparing HALO measured power to a linear power monitor photodiode scaled by \eref{eq:gain} and reported as percent relative to the HALO powers. Error bars are the expanded (95~\% confidence) uncertainty of only the HALO measurement. Discrepancies are well within the uncertainty of each measurement.}
    \label{fig:linearity}
\end{figure}

To evaluate the linearity of the HALO response to power, we compared the HALO measurements to an embedded monitor photodiode located within the laser source. The absolute power reported by the monitor is currently uncalibrated (calibration is expected to have drifted in the years since this device was last compared to a known standard with 1.2~\% expanded uncertainty); however, over the power range in question, the monitor response is linear. We start by scaling the power predicted by the photodiode monitor by the theoretical expectation of gain $G^*$ depending on the number of reflections in the HALO system and optical losses:
\begin{equation} \label{eq:gain}
    G^* = \sum_{j=1}^{j=N} (1-L)^{j-1} R^{j-1},
\end{equation}
where $N$ is either 1 or 14. Then, we plot the power measured by the HALO against the scaled monitor power and calculate a linear fit to this data. \Fref{fig:linearity} reports the residuals to this linear fit in percent relative to measured HALO powers. Error bars in this plot give the expanded uncertainty (95~\% confidence) of only the HALO measurements. Within this uncertainty, the HALO measurements of power are linear with respect to the monitor photodiode.

\subsubsection{A closer look at gain}\label{sec:gain}
\begin{figure}[t]
    \centering
    \includegraphics[width=0.4\textwidth]{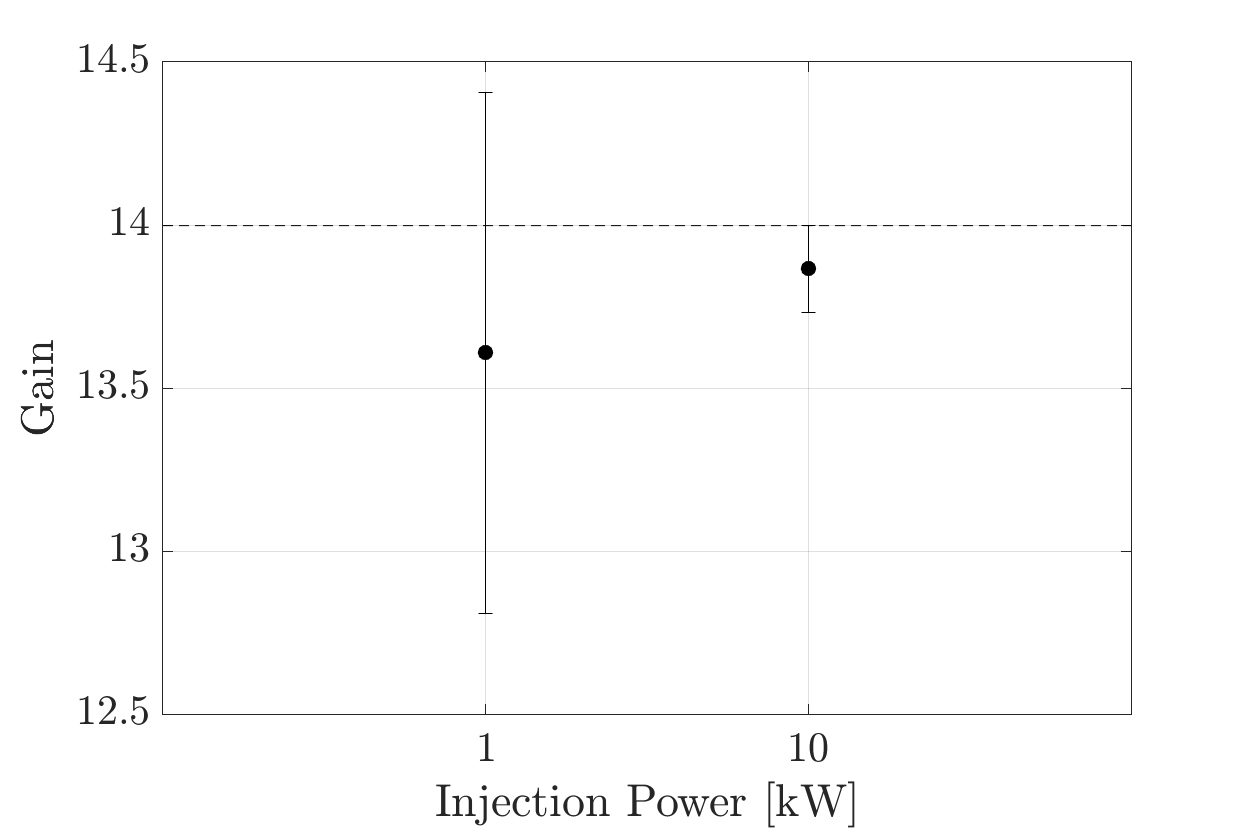}
    \caption{Measured signal gain between (ratio of) 14-reflection measurements and 1-reflection measurements by the HALO for 1~kW and 10~kW nominal input laser power. Error bars report combined, expanded (95~\% confidence) uncertainty of the 14-reflection and 1-reflection measurements. Dashed line indicates the expected gain given optical losses calculated from \eref{eq:gain}.}
    \label{fig:gain}
\end{figure}

\Eref{eq:gain} gives the theoretically expected gain of the multiple-reflection HALO system if the only losses to power amplification are optical. \Fref{fig:gain} shows the measured gain between a 14-reflection measurement and a 1-reflection measurement taken at two injected laser power levels (nominally 1~kW and 10~kW). Error bars on this plot denote the combined, expanded uncertainty (95~\% confidence) of the 14- and 1-reflection measurements; see \sref{sec:measurementUncertainty} for individual uncertainties of each measurement. The dashed line on this plot gives the theoretically expected gain factor from \eref{eq:gain}. The average gain factor measured for these power levels is consistent wih a gain factor of 14, given the measurement uncertainties. In this comparison, we rely on the monitor diode to confirm that the injected laser power was the same in the 14- and 1- reflection measurements, which is subject to significant uncertainty. Because we cannot easily improve the reference power measurement, the uncertainty of the gain linearity can only be reduced by repeating this measurement many times. Further studies will also be required to more deeply understand the influence of other loss mechanisms – like heating effects – on the expected amplification factor from the multireflection HALO system.

\subsubsection{Noise trends}\label{sec:noise}
In \sref{sec:theory}, we discuss the influence of noise and the possibility that some noise sources may be amplified along with the signal, such that our SNR improvement factor may be lower than the signal gain \eref{eq:gain}. We suspect that the dominant source of noise is turbulent air currents that disturb the position of the sensing mirror attached to the force sensor, and this source may increased with the amplified signal. We assume that air currents are largely driven by thermal gradients within and near the HALO system. Air flow from the laboratory into the 2.3~m$^3$ space within the air current damping enclosure that surrounds the HALO system is expected to be much smaller than air circulation within the enclosure. To study some of the driving factors of noise in our power measurements, we look at the variability in repeat measurements as a function of the amplified power signal, the non-amplified or injected power into the HALO system (specifically looking at the laser power incident on the beam dump as this is the hottest component within the HALO enclosure), and the intensity of laser light on the beam dump.

\begin{figure}[t]
    \centering
    \includegraphics[width=0.9\textwidth]{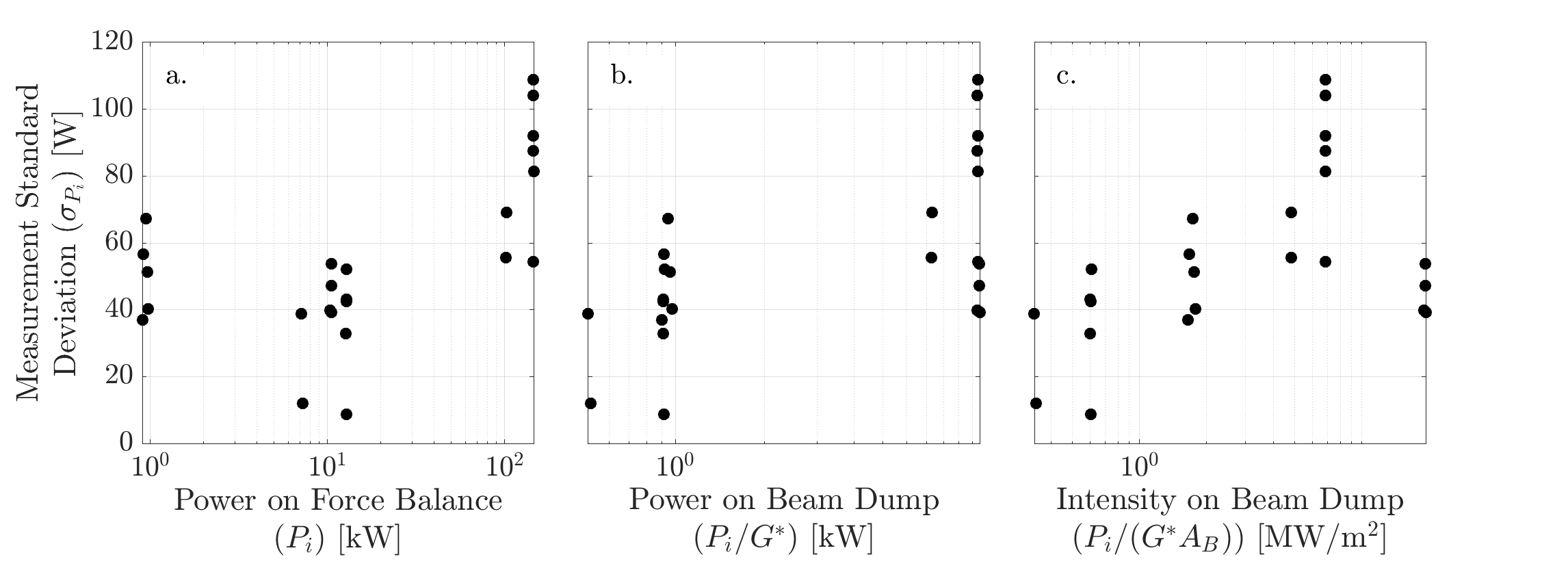}
    \caption{Measurement noise plotted against (a.) laser power on the force balance relating to $\eta_N$ in \eref{eq:noise}, (b.) laser power on beam dump relating to $\eta_P$ in \eref{eq:noise}, and (c.) laser intensity on beam dump also relating to $\eta_P$ in \eref{eq:noise}, where $A_B$ is the area of the laser in the plane of the beam dump.}
    \label{fig:noise}
\end{figure}

The estimated square root of the variance of each measurement run ($\sigma_{P_i}$) at every power level for both 14-reflection and 1-reflection measurements are used in our noise study. We provide three bases on which to view the trend of the measurement noise in \fref{fig:noise}. In \fref{fig:noise}(a), the measurement standard deviation is plotted against the power on the force sensor and relates to the $\eta_{N}$ amplification-dependent term in \eref{eq:noise}. In \fref{fig:noise}(b), this same noise is plotted against the laser power on the beam dump, labeled ``d." in \fref{fig:construction}(i.), and relates to the $\eta_{P}$ input power-dependent term in \eref{eq:noise}. \Fref{fig:noise}(c) is similar to \fref{fig:noise}(b) in that it also relates to $\eta_{P}$; however, here the measurement noise is plotted against the calculated laser intensity on the beam dump, where differences in beam size at the beam dump between the 1-reflection and 14-reflection cases are accounted for by the differences in laser propagation length in these two setups and the laser divergence angle.

The scatter of our few data points makes it difficult to attribute a clear trend to the noise in any of these three plots. However, there does seem to be a slight upward trend in all three cases, except for the highest intensity on beam dump set of data. This small trend of approximately a factor of two noise increase suggests that the noise in these measurements increases with both input laser power and total power on the force balance (or amplified power). As shown in \fref{fig:noise}(a), the measurement noise approximately doubles with an order of magnitude increase in the amplified power on the force balance. In addition, there appears to be as many as three noise regimes each defined by order of magnitude differences in the amplified laser power on the force balance - at 1~kW, 10~kW, and 100~kW, where the lowest measured noise occurs in the range of 10~kW on the force balance. The noise trends in \fref{fig:noise} appear to align with the theory that air current noise are largely driven by thermal gradients in the environment surrounding the HALO, and suggest that improvements to the system to reduce measurement noise should include further dampening of turbulent air flow in the vicinity of the force balance.

\subsection{Measurement uncertainty}\label{sec:measurementUncertainty}
We follow a sequential alignment procedure to set the position and angle of the sensing mirror and all ring mirrors of the HALO. The planes of the sensing mirror and ring plate are first set to be parallel to each other and perpendicular to gravity. Their spacing is set using an optical distance meter with 1 mm accuracy. We then use a narrow, well-collimated infrared laser to align the angle of each ring mirror such that the reflected beam is centered on the next ring mirror port. Each ring mirror is aligned sequentially until the alignment laser exits the HALO system and is incident on a beam dump. Following the system alignment of the HALO mirrors, we align our high-power infrared laser using its guide beam such that the guide beam is incident on the center of both the entrance port of the HALO and the beam dump following the HALO. Given the large parameter space of this system alignment, we developed a Monte Carlo program that propagates rays through an in-house ray tracing program to predict the distribution of measured forces given easily defined geometric bounds on each alignment step. This program is included as supplemental material. The outcome of 10,000 simulations using this sequential alignment procedure resulted in a normal distribution of forces with a relative standard deviation of 0.097~\% (distribution standard deviation divided by distribution mean times 100~\%). This is the final alignment uncertainty used in all HALO measurements where the laser is incident on the sensing mirror 14 times. In the case of a single reflection off the sensing mirror, the simulated, relative uncertainty is 0.29~\%. The relative uncertainty of the single reflection is higher than the 14-reflection setup because of the three-dimensional geometry of the HALO. Rather than alignment errors compounding after each reflection, these random errors average out with increased number of reflections because the direction of misalignment rotates as the laser travels through the system.

\begin{figure}[b]
    \centering
    \includegraphics[width=0.6\textwidth]{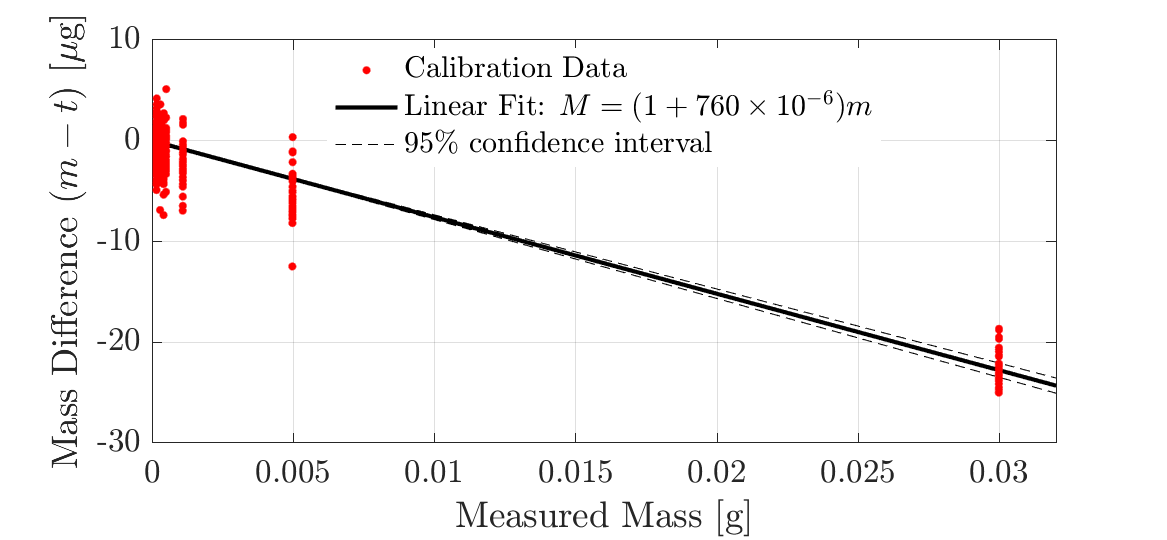}
    \caption{Commercial force balance calibration curve with linear fit giving the calibration factor from measured mass in grams ($m$) to true mass as given by prior artifact calibration ($M$). The relative uncertainty in this linear calibration factor is $0.0012$~\%.}
    \label{fig:scaleCalib}
\end{figure}

Calibration of the commercial force balance is performed prior to any laser power measurements by repeatedly recording the balance response to a set of NIST calibrated mass artifacts from 160~$\mu$g to 30 mg with $0.0005$~\% uncertainty \cite{Shaw2019b, Rogers2020}. A linear fit between the calibrated mass of each artifact and the measured mass from the commercial force balance sets the calibration factor, which, for the balance used in these HALO measurements, is $M = (1+760\times10^{-6})m$, where $M$ is the calibrated mass of each artifact and $m$ is the measured mass of each artifact as given by the commercial force balance, \fref{fig:scaleCalib}. By standard linear regression, the relative uncertainty of this calibration factor is $0.0012$~\%. Because the force balance is calibrated prior to its integration into the HALO system, we must add an additional uncertainty to this calibration of $0.00007$~\% for the 1~mrad precision in aligning the force balance to gravity in both the calibration setup and in the HALO. By adding in quadrature the artifact mass uncertainty, the calibration factor uncertainty, and the gravity alignment uncertainty, we obtain a total relative force balance calibration uncertainty of $u_{cal} = 0.0013$~\%.

The final combined expanded relative uncertainty in the HALO laser power measurements is defined by
\begin{equation} \label{eq:finalUnc}
    U_C = \mathrm{k}\sqrt{u_{cal}^2+u_a^2+u_R^2+\left(\frac{\sigma_{\bar{P}}}{\bar{P}}\right)^2},
\end{equation}
where a coverage factor of k is applied to obtain 95~\% confidence bounds. All of the uncertainty components, including the coverage factor for each measurement set, are defined in \tref{tbl:unc14} for the 14-reflection measurements and \tref{tbl:unc1} for the 1-reflection measurements at the two main power levels of interest - 10~kW and 1~kW. When the full HALO system is utilized with 14 laser reflections off the sensing mirror, the relative expanded uncertainty of a nominally 10~kW input power measurement is 0.26~\% and is 0.46~\% for a 1~kW input power measurement. This compares to the measurements using only a single laser reflection off the sensing mirror having expanded uncertainties of 1.14~\% and 6.9~\% for a 10~kW and 1~kW input power level, respectively. This is better than a factor of 4 improvement in relative uncertainty at 10 kW and about a factor of 15 improvement at 1~kW. 

\begin{table}[hb]
    \caption{\label{tbl:unc14}Uncertainty Budget for the 14-reflection measurements with type (A-measured or B-estimated) of uncertainty and distribution of each uncertainty source. The final combined uncertainty is given with a coverage factor (k) that reflects the degrees of freedom in that measurement.}
    \begin{tabular}{@{}llll}
        \br
        Uncertainty Component & Type & Distribution & Relative Uncertainty  \\
        \mr
        Force balance ($u_{cal}$) & A & normal & 0.0013~\% \\
        Alignment ($u_a$) & B & normal & 0.097~\% \\
        Mirror reflectance ($u_R$) & B & rectangular & 0.0006~\% \\ 
        Measurement variability $\left(\sigma_{\bar{P}}/\bar{P}\right)$ & A & normal &  \\
        \hspace{1em}10~kW (145.65~kW) & & & 0.0252~\% \\
        \hspace{1em}1~kW (12.79~kW) & & & 0.136~\% \\
        \mr
        \multicolumn{4}{l}{Combined Expanded Uncertainty ($U_{C}$) (Eq.~\ref{eq:finalUnc}) (95~\% confidence)}  \\
        \multicolumn{2}{l}{\hspace{1em}10~kW (145.65~kW)} & k=2.57 & 0.26~\% \\
        \multicolumn{2}{l}{\hspace{1em}1~kW (12.79~kW)} & k=2.78 & 0.46~\% \\
        \br
    \end{tabular}
\end{table}

\begin{table}[ht]
    \caption{\label{tbl:unc1}Uncertainty Budget for the 1-reflection measurements with type (A-measured or B-estimated) of uncertainty and distribution of each uncertainty source. The final combined uncertainty is given with a coverage factor (k) that reflects the degrees of freedom in that measurement.}
    \begin{tabular}{@{}llll}
        \br
        Uncertainty Component & Type & Distribution & Relative Uncertainty  \\
        \mr
        Force balance ($u_{cal}$) & A & normal & 0.0013~\% \\
        Alignment ($u_a$) & B & normal & 0.29~\% \\
        Mirror reflectance ($u_R$) & B & rectangular & 0.0006~\% \\ 
        Measurement variability $\left(\sigma_{\bar{P}}/\bar{P}\right)$ & A & normal &  \\
        \hspace{1em}10~kW (10.51~kW) & & & 0.0216~\% \\
        \hspace{1em}1~kW (0.94~kW) & & & 2.46~\% \\
        \mr
        \multicolumn{4}{l}{Combined Expanded Uncertainty ($U_{C}$) (Eq.~\ref{eq:finalUnc}) (95~\% confidence)}  \\
        \multicolumn{2}{l}{\hspace{1em}10~kW (10.51~kW)} & k=3.18 & 1.14~\% \\
        \multicolumn{2}{l}{\hspace{1em}1~kW (0.94~kW)} & k=2.78 & 6.9~\% \\
        \br
    \end{tabular}
\end{table}

\section{Discussion}\label{sec:discussion}
Our measurements demonstrate a favorable reduction in laser power measurement uncertainty by incoherent amplification of laser pressure in a radiation pressure-based system. In theory, this technique should improve the SNR of a laser power measurement by a factor very close to the number of times the laser reflects off the sensing mirror. However, we do see some additional amplification of noise that has an impact on the uncertainties we measure. 

At the highest power levels measured, we observe an increase in the measurement noise as given by the square root of the estimated variance in a set of 10 sequential laser injections ($\sigma_{P_i}$). Though this trend is not clear due to the large spread in the measured noise from repeat independent measurements, the slight upward trend with increased power both on the force balance, representing the amplified power, and on the beam dump, representing the pre-amplified input power, suggests that thermally driven effects contribute to increased noise. We know from prior tests that the dominant noise in the force balance measurement comes from air currents that disturb the position of the sensing mirror affixed to the force balance. The sensing mirror, with its large 177~cm$^2$ area acts as a sail, catching the small air currents. These disturbances can be large relative to the power signals being measured. As such, improvements to the uncertainty of these measurements should involve reduction of air currents near the force balance, and, in particular, should take into account thermally driven air currents as the input laser power is increased and the force balance is exposed to larger amounts of optical power both directly on the mirror and from scatter within the optical system.


Our sequential alignment procedure in these measurements is not the optimal method for minimizing alignment uncertainties. As the number of reflections increases, errors in alignment of each reflection increases due to the small, but significant divergence of the alignment laser. This results in larger force uncertainty contributions assigned to the later reflections. Some of the added uncertainty is overcome by the amplification of the force signal and averaging of random errors from each reflection (due to the three-dimensional geometry); however, an optimal alignment procedure would not produce an uncertainty dependent on the number of reflections. The spot pattern of laser incidence locations on the sensing mirror is extremely well defined from the geometry of the system and may be used to refine the angular alignment of each ring mirror independently. The HALO mirrors can also be referenced to a static target that is calibrated separately to decouple the alignment of each mirror from the others. Our ray-based, Monte Carlo simulation of misaligned force measurements predicts a normal distribution of alignment errors. Therefore, another method to minimize the contribution of alignment uncertainties is to realign the HALO system between repeated measurements to minimize this uncertainty source through averaging. In this report, our estimate of the measurement uncertainty resulting from alignment errors of the HALO mirrors is based entirely on the Monte Carlo and ray tracer simulation of the alignment geometry. Future experimental work will demonstrate its validity.

Our tests of the HALO multiple reflection system take advantage of a convenient commercial force balance that requires prior calibration to provide a mass traceable measurement of laser power. A preferred method of measuring the radiation pressure force of the laser is to use an electrostatic force balance (EFB) \cite{Shaw2019b,Keck2021}. These force sensors not only are highly sensitive to small forces, they also intrinsically deliver a force directly traceable to electrical units, obviating the need for calibration of the sensor response to mass artifacts. 

A dedicated EFB is currently under development for integration with the HALO system \cite{Keck2021}. If we look at the combined HALO measurement uncertainty \eref{eq:finalUnc}, use of the SI-traceable EFB in place of the calibrated force balance will substitute the force balance calibration uncertainty term ($\sigma_{cal}$) with what is expected to be a much smaller uncertainty on the EFB itself. In addition to limiting the measurement uncertainty of the HALO, calibration of the commercial force balance to mass artifacts provides a weaker link to fundamental units than an EFB standard will. One will note that our HALO measurements herein are not validated against any other standard reference. This is because, at these high-power levels, no other power measurement system is capable of discerning laser power to better than 1-2~\% \cite{Williams2019}. To validate the low measurement uncertainty of the HALO system, we must compare to force standards rather than other laser power standards and the EFB will provide the best reference in this validation test.

\section{Conclusion}\label{sec:conclusion}
This report documents the first measurements of laser power from 500~W to 10~kW with a radiation pressure measurement system where the laser pressure signal is amplified incoherently 14 times. We assess the force measurement with this system, deemed the High Amplification Laser-pressure Optic (HALO). In addition to outlining the amplification theory, we examine measurements of laser power with the HALO system and confirm that these measurements approximately agree with expected levels of gain; though, some thermally driven deviations from expectation are denoted. Four primary sources of uncertainty in the power measurement are defined. The largest sources of uncertainty in the report are the measurement noise and alignment of the optical elements of the HALO. For an input laser power of nominally 10~kW, we measure a total expanded (95~\% confidence) uncertainty of 0.26~\%, and for a nominal input power of 1~kW we measure an expanded relative uncertainty of 0.46~\%. At these power levels, prior state-of-the-art absolute power measurements report uncertainties on the order of 1-2~\%. Our present measurements with the HALO system represent an improvement in measurement uncertainty by roughly a factor of four. Planned advancements to the optical alignment, suppression of heat-driven noise, and changing of the force balance from one requiring mass artifact calibration to a superior electrostatic force balance will further decrease the uncertainty of these high-power laser measurements and provide a more direct link to SI units, improving the confidence in the absolute scale of these measurements. This report is the first phase of a larger effort to reach relative uncertainty levels comparable to that of the cryogenic radiometer (0.01~\%) at power levels a billion times higher in magnitude.


\ack
We thank Gordon Shaw, David Newell, Dmitry Vorobiev, and Paritosh Manurkar for valuable discussions relating to the mass calibration of the commercial scale, insights into noise and representations of our noise measurements, and for critical review of this manuscript.


\section*{References}
\bibliographystyle{unsrt}

\end{document}